# Shake and sink: liquefaction without pressurization

C. Clément[1], R. Toussaint*[1], E. Aharanov[2]

[1]University of Strasbourg, Institut de Physique du Globe de Strasbourg, CNRS UMR7516, 5 rue Descartes, F-67000 Strasbourg, France

[2]Institute of Earth Sciences, The Hebrew University, Jerusalem, 91904, Israel

Corresponding author: Renaud Toussaint (renaud.toussaint@unistra.fr)

**Key Points:**

- A mechanism for soil liquefaction during earthquakes, which does not require elevated pore pressure, but may be enhanced by it.
- Offer a theory for seismic liquefaction controlled by buoyancy in saturated soils. Predictions confirmed by experiments & simulations.
- Explanation of previously puzzling far-field and repeated liquefaction occurrences and liquefaction under drained conditions.

**Abstract**

Soil liquefaction is a significant natural hazard associated with earthquakes. Some of its devastating effects include tilting and sinking of buildings and bridges, and destruction of pipelines. Conventional geotechnical engineering practice assumes liquefaction occurs via shear-driven compaction and consequent elevation of pore pressure. This assumption guides construction for seismically hazardous locations, yet evidence suggests that liquefaction strikes also under currently unpredicted conditions. Here we show, using theory, simulations and experiments, another mechanism for liquefaction in saturated soils, *without necessitating high pore fluid pressure* or *special soils*, whereby seismically triggered liquefaction is controlled by buoyancy forces. This new mechanism supplements the conventional pore pressure mechanism, enlarges the window of conditions under which liquefaction is predicted to occur, and may explain previously not understood cases such as liquefaction in well-compacted soils, under drained conditions, repeated liquefaction cases, and the basics of sinking in quicksand. These results may greatly impact hazard assessment and mitigation in seismically active areas.

# 1 Introduction

Liquefaction occurs in saturated soils when an initially rigid soil, which supports structures, changes rheology under earthquake-induced shaking [*Diaz-Rodriguez et al,* 1992; *Wang et al,* 2010; *Youd et al,* 1978] to a liquid-like slurry, in which structures such as buildings sink and tilt [*O'Rourke et al,* 1989] and structures such as pipelines float [*O'Rourke et al,* 1989; *Ambraseys et al,* 1969; *Huang et al,* 2013]. Liquefaction is of crucial importance in geoengineering [*Youd et al,* 1978; *Youd et al,* 2001; *Hausler et al,* 2001; *Seed et al,* 2003; *Sawicki et al,* 2009], and constitutes a basic physics question for the two-phase system of fluids and grains [*Khaldoun et al,* 2005; *Brzinski et al,* 2013; *Huerta et al,* 2005; *Lohse et al,* 2004].

The *conventional mechanism* for explaining liquefaction requires *un-drained* and *under-compacted* saturated soils. It assumes that during earthquakes the induced cyclic shear causes pore structure in under-compacted porous soils to collapse. The un-drained condition leads to trapping of fluid in compacting pores, so pore pressure increases until its value may approach the total stress [*Youd et al,* 2001]. High pore pressure values lead to loss of strength and liquefaction of soils. Indeed, this mechanism was confirmed in many liquefaction instances [e.g. *Obermeier et al,* 1996; *Holzer et al,* 1989], yet it fails to predict many other observed situations:

1. **Liquefaction in pre-compacted soils.** An example of such a scenario is the liquefaction event in Kobe, Japan, following the 1995 Great Hanshin Earthquake (M=6.9). *Soga* [1998] reviewed the damage in the port facilities that were built on a reclaimed island. It was found that soils that were vibro-compacted, and therefore are not expected to be compactive, were still liquefied. This occurrence is baffling if liquefaction is explained solely by a mechanism that involves compaction of initially loose soils.

2. **Recurrent liquefaction events** [*e.g. Wakamatsu et al,* 2012; *Ambraseys et al,* 1969; *Youd et al,* 1978]. As explained by *Obermeier* (1996): "liquefaction has a strong tendency to recur at the same site" but "An apparent contradiction to recurrent liquefaction at the same site is the observation that liquefaction commonly densifies sediments. Densification should reduce the liquefaction susceptibility. Worldwide engineering measurements before and after occurrences of liquefaction indicate that thick zones … densified substantially whenever liquefaction was severe [*Koizumi,* 1966*; Ohsaki,* 1970]." Thus recurrent liquefaction is not predicted by the conventional mechanism.

3. **Far-field liquefaction that occurred despite small seismic energy input.** Experiments find that liquefaction via the conventional mechanism requires a minimum energy density input: at least 30 J/m$^3$. However, about half of the sites that underwent liquefaction during earthquakes received less energy than that, probably by orders of magnitude less [*Wang,* 2007; *Manga et al,* 2012].

4. **Liquefaction produced under drained conditions** [*Goren et al,* 2010; *Goren et al,* 2011; *Lakeland et al,* 2014]. Although the conventional mechanism explains liquefaction by pore pressure increase in rapidly compacting soils, most liquefaction demonstrations actually show fully drained liquefaction and involve no pore pressure rise. See e.g. the demo of Illinois Geological Survey https://www.youtube.com/watch?v=cONq231dn6w

The above observations are important in that they are both widespread and fundamentally inconsistent with the conventional mechanism of liquefaction. This paper presents an alternative liquefaction mechanism that requires neither compactive soils, nor high pore pressure. The new mechanism occurs in saturated soils. It is triggered by seismic accelerations and controlled by buoyancy effects. Although in nature liquefaction may occur solely by the mechanism we propose, we show below that the new mechanism may also combine with, and be enhanced by, the *conventional mechanism,* i.e. by elevated pore pressure.

In what follows we present experiments, theory and simulations. All of those investigate a system similar to the common table-top systems used to demonstrate the process of liquefaction to students, (see point 4 above). Our model system comprises an intruder, placed on top of a partially or fully water-saturated granular layer. The system is shaken at specified accelerations and amplitudes to simulate a building or a structure experiencing seismic shaking. Under a certain range of shaking conditions, the medium liquefies and the intruder sinks.

We define liquefaction based on measurements of the intruder sinking. This definition is similar to that used in the field, where the occurrence of liquefaction is identified based on the phenomenology typical of liquefaction, such as sinking and floating of structures. We do not define here liquefaction by its narrower mechanistic definition (sometimes used by engineers) which identifies liquefaction with high pore pressure, as the mechanistic definition is restrictive and does not capture the full range of occurrences, which is what really matters when we wish to compare theory with field observations, as is done here.

Our three method of analysis, experiments, theory and simulations, all show remarkable agreement, allowing us to derive a phase diagram which maps the conditions for liquefaction occurrence by the newly identified mechanism. Using this analysis we predict the conditions under which a granular saturated soil can liquefy under the new mechanism, as function of the amplitude and frequency of the shaking. We apply this analysis to evaluate liquefaction potential of natural sites as function of their distance from Earthquakes of a specific magnitude, and show that the proposed mechanism may explain many previously unexplained natural liquefaction events.

## 2 Experimental Methods

The experiments comprise a square transparent glass box, 10 cm each side, linked to a step motor and a shaker. The motor (produced by Phidgets) is 1063 PhidgetStepper Bipolar 1- for movements at small frequency and high amplitude, and the shaker (produced by Tira) is S 51120 with the amplifier BAA500 - for high frequency and small amplitude. The box is filled with monodisperse (1% dispersion in diameters) Ugelstad beads of diameter 140 μm (Dynoseeds TS 140-51) and density 1.05 kg/dm$^3$. We estimated the static friction coefficient of the material at 0.48 by measuring the angle for sliding initiation of a thick layer. For dry media the beads are directly poured into the box. For wet media the box is first filled with a prescribed amount of water then we gently drop the beads as rainfall to avoid trapping air bubbles. To obtain a controlled and reproducible initial compaction we systematically uncompact the medium with a knife by applying few shearing movements. Subsequently, the box is shaken at 10Hz at amplitude of 1cm during one minute. Eventually we let the medium settle for three minutes. Hence the medium is assumed to be a close random pack with a porosity around 0.37. After this resting time a larger sphere, termed "the intruder", is positioned on the surface of the medium. The intruder is printed out of ABS plastic material using a MakerBot Replicator2X. In the following experiments, we use an intruder of density 1.03 kg/dm$^3$. After positioning the intruder at the surface of the granular material, we wait one minute for the system to equilibrate, and start the shaking engine. The box is shaken horizontally under a given acceleration ($a \in$ [10$^{-2}$; 100] m/s², $\Gamma = a/g$=0.001 to 10 with $g$=9.81 m/s²) and frequency ($f \in$ [0.15; 50] Hz), corresponding to conditions typically met in earthquakes with macroseismic intensity II to V-VI [*Souriau,* 2006].

All experiments are recorded by a Nikon 5200 camera at 25 frames per second. Thresholding the pixel colormap values of the pictures, we extract the height of the intruder and the height of the medium surface. Both are corrected from the view angle and the emerged height of the intruder is computed at each time. From the emerged height we compute the associated emerged volume $V_{em}$ and immersed volume $V_{im}$ of the intruder.

## 3 Experimental Results

### 3.1 Experimental Results: Liquefaction revealed through the sinking of an object.

In order to test the physics of the process, experiments were ran using 3 different water-level conditions: *dry (no added water), saturated* (where water fills the granular media, and their level reaches the ground level) and *fully-immersed*, (where water level is above the intruder top) (Figs 1, 2a). Videos of the experiments can be found in the Suppl. Material.

Figure 1 compares the sinking of intruders during shaking experiments performed under the 3 different conditions: Under dry conditions no sinking occurs (Fig 1d). Under saturated conditions, the intruder remains at the surface for low-acceleration shaking (Fig 1b) but sinks and reaches a steady state when subjected to larger accelerations (Fig 1a). For fully immersed conditions (Fig 1c), a slight sinking occurs but remains 3 times smaller than for saturated cases. Figure (1e) quantifies the normalized emerged volume $\Sigma_1(t) = \frac{V_{em}(0) - V_{em}(t)}{V_{TOT}}$ for each case. Here $V_{em}(0)$ is the emerged volume before initiation of shaking (the volume of the intruder above the ground level), $V_{em}(t)$ is the emerged volume at time t during shaking, and $V_{TOT}$ is the total volume of the intruder.

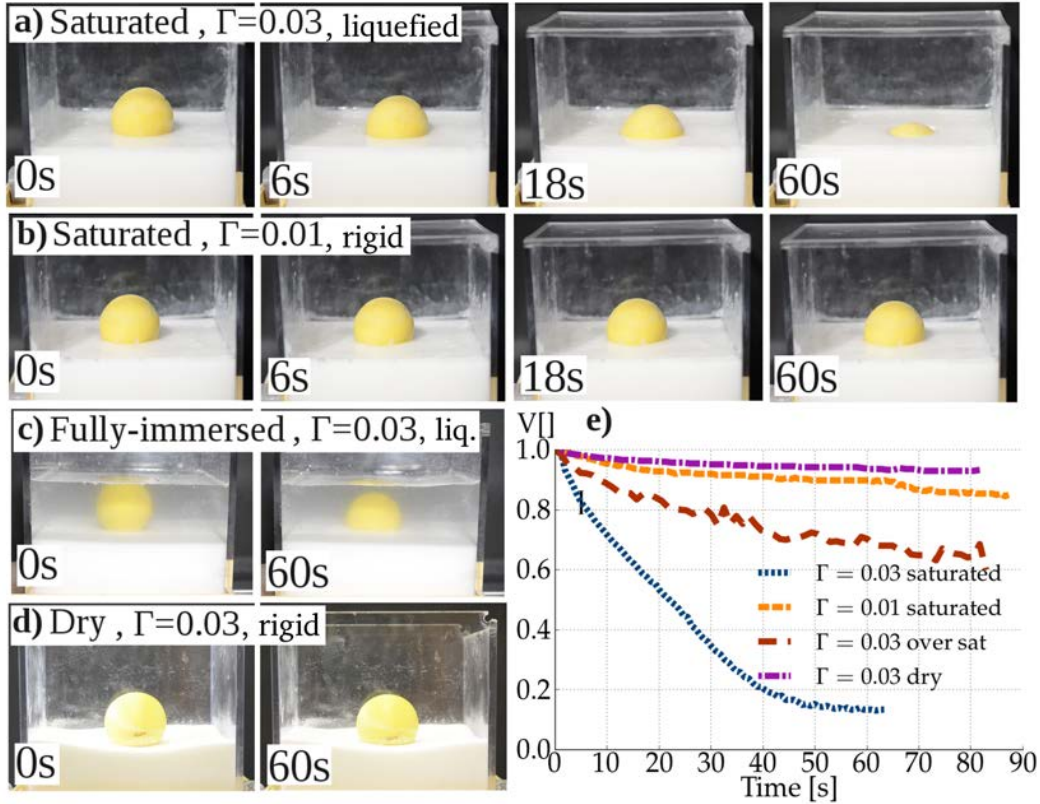

Figure 1: Snapshots from experimental shaking of wet granular media show that intruder sinking depends on fluid levels and acceleration conditions: a,b) Saturated runs: the intruder sinks (case (a)) and continuously progresses towards a new steady-state, for large Γ. The intruder does not sink for small Γ (case (b)). c) In contrast to case (a), fully immersed run shows only slight sinking under large Γ. d) Dry run, showing no sinking. e) Normalized emerged volume $\Sigma_1(t)$ of the intruder as function of shaking time, in the four cases, shows largest sinking occurs at saturated case and high acceleration. Videos: see Suppl. Movie S1 (a), Movie S2 (b), Movie S3(c), Movie S4 (d)

We define the system as "liquefied" when large sinking occurs, as in Fig 1(a, c), and as "rigid" when small or no sinking occurs as in Figs 1(b, d) (for practical experimental definitions as function of thresholds see section 6). Results show that liquefaction occurs in saturated and fully immersed conditions when shaking acceleration exceeds a critical value (Fig 1e).

### 3.2 Theoretical key point: Liquefaction occurs without elevation of the pore pressure.

In this subsection we show that pore pressure did not rise during our experiments. To predict pore pressure (PP) in our experiments we follow equation (31) in *Goren et al,* [2011], which define a non-dimensional Deborah number, $De = t_d/t_0$, measuring the ratio of times to relax PP by diffusion $t_d$ versus the time to build PP, $t_0$, during a specific soil deformation process.

When $De < 1$ the system is well drained, while when $De > 1$ PP is expected to build up [*Goren et al,* 2011; *Lakeland et al,* 2014; *Goren et al,* 2010]. We associate $t_0$ with the period of the shaking, ranging in our experiments between 0.15-50 Hz, so $t_0$ is 0.02 - 10s. The time for dissipation $t_d$ is expressed [*Goren et al,* 2011] in terms of *l,* the depth of the layer which liquefied, and *D,* the PP diffusion coefficient, as $t_d = l^2/D$. The diffusion coefficient is equal to $D = K/\beta\eta\Phi$ where *K* is the permeability, $\Phi$ is porosity, $\eta$ fluid viscosity and $\beta$ fluid compressibility. Our experimental box has $l = 10 cm$, $\Phi > 0.36$, $K > 10^{-11}$ m² [*Carman,* 1937], $\beta = 5.10^{-10}$ Pa⁻¹ and $\eta = 1.10^{-3}$ Pa.s (compressibility and viscosity of water), thus $t_d < 0.0002$ s, and in our experiments $De<<1$, the system is well drained, and there is no possibility of pore pressure buildup.

## 4 Theoretical analysis of the proposed seismic liquefaction mechanism: liquefaction controlled by buoyancy.

### 4.1 Intruder sinking is triggered by external accelerations and affected by buoyancy.

The physics behind the phenomena in Fig 1 is straightforward: Fig 2 shows a schematic cartoon of a shaken granular medium, with an intruder on top. We consider a saturated granular medium of grain density $\rho_g$ and two grains of this medium, *i* and *j* (Fig 2b). The intruder, which is a big grain, called B, of density $\rho_B$ and volume $V_{TOT}$, is positioned on the surface. The system is in mechanical equilibrium before shaking and then undergoes horizontal motion of the form $A \sin(\omega t)$ with a peak acceleration $A\omega^2$ and frequency $f = \omega/2\pi$. We seek conditions under which the system's components will start to rearrange. If the sum of the external forces is equal to zero, the medium will move as a rigid block in the reference frame of the moving media. If not, there will be accelerations between grains and the system will rearrange. We next compute the mechanical forces between the different grains.

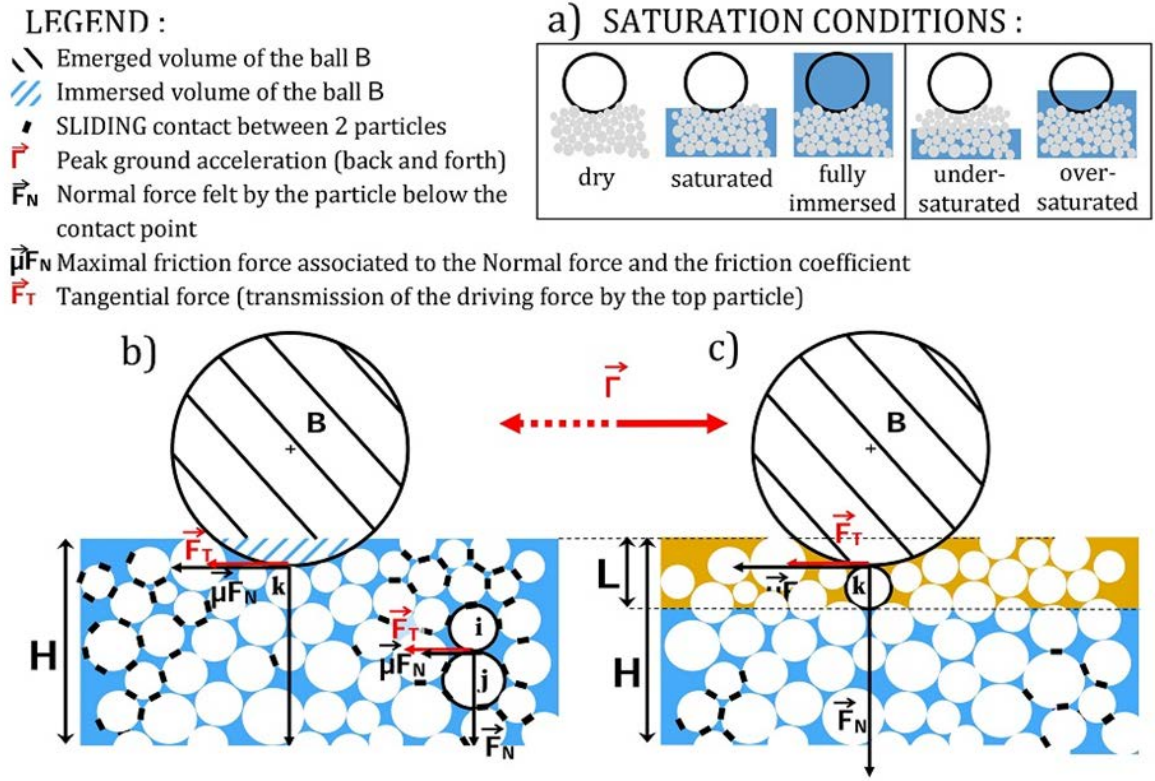

Fig 2: Conceptual model of the physics by which differential buoyancy controls seismically triggered liquefaction. a) The different saturation conditions used in experiments, simulations, and theoretical analysis. b) Liquefaction process of saturated medium: When shaken with $\Gamma$ **above a critical** $\Gamma_L$ contacts between immersed particles away from the intruder are able to slide (black points), while particle contacts in the region below the intruder can't slide. The sliding medium rearranges everywhere except below the intruder, acting as an effective liquid in which the intruder sinks to its isostatic position. c): Theory predicts that when water doesn't reach the ground level, leaving a dry zone of thickness *L*, liquefaction will start at depth.

First we focus on contact *i-j* between particles *i* and *j* (Fig 2b). These two grains are entirely immersed. The normal force acting on the bottom grain *j* is the sum of the weight of the overlying column plus the buoyancy exerted over this column: $\vec{F}_{ij}^n = m_{above}(1 - \rho_w/\rho_g)\vec{g}$, where $m_{above}$ is the mass of the overlying grain-column, $\rho_w$ is water density and $\vec{g}$ is gravitational acceleration. Assuming horizontal motion of the form $A\sin(\omega t)$, mechanical equilibrium of the column, and negligible lateral stress transfer, contact *ij* experiences a tangential force $F_{ij}^t = A\omega^2 \, m_{above}$. If the peak ground acceleration $A\omega^2$ is large enough, $F_{ij}^t$ will exceed the threshold set by the Coulomb criterion $\mu \, F_{ij}^n$, where $\mu$ is the friction coefficient, and contact *ij* will slide (we neglect horizontal stress gradients, because the whole layer starts to slide). If the inertial lateral force induced by shaking is below the frictional threshold $\mu \, F_{ij}^n$, the granular medium moves as a rigid block. The normalized acceleration $\Gamma = \frac{A\omega^2}{g}$, above which immersed grains in the medium will slide past each other, is $\Gamma_L = \mu(\rho_g - \rho_w)/\rho_g$.

Next we focus on the contact between grain B and the particle below it, called *k* (Fig 2b). If B is entirely emerged (i.e. no part of it is sunk in water), the normal force it exerts on *k* is $\vec{F}_{Bk}^n = V_{TOT}\rho_B\vec{g}$. According to a Coulomb criterion, if the inertial lateral force induced by shaking, $F_{Bk}^t = A\omega^2 V_{TOT}\rho_B$ exceeds $\mu F_{Bk}^n$, the intruder will slide on particle *k*, i.e. if $\Gamma > \mu = \Gamma_{GE}$ the intruder slides.

We therefore predict three regimes of behavior according to the applied acceleration:

$\Gamma < \Gamma_L = \mu(\rho_g - \rho_w)/\rho_g$     ***Rigid***     (1a)

$\Gamma_L < \Gamma < \Gamma_{GE} = \mu$     ***Heterogeneous Liquefaction, H.L***     (1b)

$\Gamma_{GE} < \Gamma$     ***Global Excitation Liquefaction, G.E.L***     (1c)

In the ***rigid*** case, equation (1a), the shaken system moves as a rigid body, since $\Gamma$ isn't sufficient to induce sliding on any contact. In the ***G.E.L*** case, equation (1c), $\Gamma$ is large enough so both grain-grain contacts (e.g. contact *i-j*) and intruder-grain contacts (e.g. contact *k*-B) slide, and the whole medium rearranges. In the intermediate case, equation (1b), termed ***Heterogeneous Liquefaction***, grain-grain contacts can slide, but B can't.

### 4.2 The equilibrium position of the intruder

In both types of liquefaction, **H.L.** and **G.E.L**, the intruder, B, sinks as the medium rearranges around it, progressively approaching a new steady-state depth, defined by its isostatic position inside a liquid medium of effective density: $\rho_{eff} = \rho_w \Phi + \rho_g(1-\Phi)$, where $\Phi$ is the porosity of the granular medium. If $\rho_{eff} < \rho_B$, the intruder will become fully immersed in the medium. If $\rho_B < \rho_{eff}$ it will end up partially immersed, with an immersed volume $V_{imISO} = V_{TOT} \frac{\rho_B}{\rho_{eff}}$.

When the top of grains and top of water do not coincide, the predicted equilibrium depends also on the volume immersed in water, $V_{im\_water}$, and that immersed in the granular medium, $V_{im_{granular}}$: $V_{im\_water}\rho_w + V_{im\_granular}\rho_g(1-\Phi) = V_{TOT}\rho_B$.

In the fully immersed condition $V_{im\_water} = V_{TOT}$ (see Fig. 2a) so the equilibrium in the fully immersed condition is $V_{im\_granular} = V_{TOT} \frac{\rho_B - \rho_w}{\rho_g(1-\Phi)}$. In the limit of $\rho_B = \rho_w$, equilibrium dictates $V_{im\_granular} = 0$ for the fully immersed case. Since our intruder density is only 3% heavier than water, this explains the negligible sinking observed in Figs. 1c and 3a.

### 4.3 The initial state of the system effects the onset of *globally excited liquefaction*

The initial immersed volume of the intruder $V_{im}(0)$ controls the acceleration threshold for *globally excited liquefaction (G.E.L)*. Indeed if the intruder is partially immersed inside water or inside the saturated granular medium before shaking, it will experience buoyancy because of the presence of water and $\Gamma_{GE}$ needs to be corrected. In these cases, the normal force B exerts on $k$ is $\vec{F}_{Bk}^n = (V_{TOT}\rho_B - V_{im}(0)\rho_w)\vec{g}$. The threshold at which B will start to slide on $k$ must be corrected:

$$\Gamma_{GE} = \mu(1 - \frac{V_{im}(0)\rho_w}{V_{TOT}\rho_B}) \qquad (2)$$

This equation corresponds to equation (1) corrected by the buoyancy force applied to the immersed part of B. In this case the G.E.L behavior is reached at a lower acceleration than if the intruder has no immersed volume. This equation allows us to make accurate predictions for transition from heterogeneous liquefaction to G.E.L. When the intruder is entirely immersed in water, equation (2) predicts $\Gamma_{GE} = \mu\left(1 - \frac{\rho_w}{\rho_B}\right) = \Gamma_L$, which means that the window between H.L. and G.E.L. disappears for fully immersed objects. In this case any acceleration greater than $\Gamma_L$ will generate sliding both among the particles of the medium and below the intruder.

### 4.4. Predicting buoyancy-controlled liquefaction under different water levels

The previous subsection calculated the reduction of $\Gamma_{GE}$ due to initial partial immersion of the intruder in water. In this subsection, we will consider a general case where water level does not necessarily coincide with the top of the granular layer, and we shall calculate how $\Gamma_L$ changes.

For an under-saturated or over-saturated medium (Fig 2a) we can split the soil into two layers. The top of the water level does not necessarily coincide with the ground level, and we consider a height difference $L$ between both – i.e. for over-saturated media, a clear water layer of depth $L$ overlays the saturated grains, and for under-saturated medium, a layer of dry soil of thickness $L$ separates the top of the grains and the water-table (Fig 2c). $\Phi$ is soil porosity. At depth $H$ under the ground level, the effective normal stress is $S^n = [H(1-\Phi)\rho_g - (H-L)(1-\Phi)\rho_w]g$ for an under-saturated medium and $S^n = H(1-\Phi)(\rho_g - \rho_w)g$ for an over-saturated medium. If the system undergoes horizontal motion of the form $A\sin(\omega t)$, the condition for sliding of grains inside the medium is $\mu S^n < (1-\Phi)H\rho_g A\omega^2$. This leads to the following criteria for liquefaction onset:

$$\Gamma > \Gamma_L = \mu\left(\frac{(\rho_g-\rho_w)}{\rho_g} + \frac{L\rho_w}{H\rho_g}\right) \qquad (3a)$$

in the case of a under-saturated medium, and to:

$$\Gamma > \Gamma_L = \mu\frac{(\rho_g-\rho_w)}{\rho_g} \qquad (3b)$$

for an over-saturated medium.

While the condition for liquefaction onset is the same for saturated media, equation (1a) and for over-saturated media, equation (3b), and is depth independent, in an under-saturated medium a larger acceleration is needed for liquefaction onset (equation (3a)). Also the depth at which liquefaction will initiate is different for different saturations: In the saturated and oversaturated cases the first soil layer to detach will be the one with the smallest friction, μ. In contrast, equation (3a) predicts that for an under-saturated medium, with an overlying dry layer $L$, the first layer to liquefy will be deep, at large $H$, since as the ratio $L/H$ decreases, the acceleration needed for liquefaction decreases (in the case of homogeneous friction threshold μ). This is consistent with recent observations of the depth of sandy soil layers that liquefied during earthquakes, up to 20m during 2008 Wenchuan [*Yuan et al,* 2009] and 12-16m for 2011 Tohoku [*Bhattacharya et al,* 2011]. This prediction may also explain cases where water appears on the surface of liquefied soils with a delay after shaking, as during the spectacular liquefaction following 1964 Niigata earthquake where sandblows occurred after three minutes of shaking [*Kawakami et al,* 1966; *Ambraseys et al,* 1969]. The delay is explained here by the fact that liquefaction will start at depth, where $L/H$ is small, and then water will need time travel upward.

In the case of a medium with non-homogeneous friction coefficient µ(*H*), liquefaction will first occur at a depth *H* that minimizes $\mu(H)\left(\frac{(\rho_g - \rho_w)}{\rho_g} + \frac{L\rho_w}{H\rho_g}\right)$, i.e. for soil layers of large *H* (deep ones) and of a small friction coefficient µ(*H*). If the thickness of the dry soil layer, *L*, is reduced to 0, $\Gamma_L$ reduces to the expressions established in equations (1a-1c), which provides also the lowest liquefaction thresholds. This also agrees with the fact that quicksands are typically saturated up to the surface. The Mt Saint Michel bay (France) counts numerous quicksands where the beaches are saturated in water [*Lefeuvre et al,* 2002], but the higher lands are safe although they are composed of the same sediments, but there the water table doesn't reach ground level.

**5 Numerical simulations of the experimental situations**

In order to test the proposed mechanism, we devise simulations that include only buoyancy and accelerations, and no pore pressure apart from the one due to gravity, i.e. a hydrostatic profile from the top of the waterbed. The simulations demonstrate that intruder sinking and medium liquefaction occur without addition of extra pore pressure.

**5.1 simulation conditions**

Simulations used a 2D Discrete Elements Model (DEM) code [*Cundall et al,* 1979], adding water-induced buoyancy forces to grains and intruder in the saturated zone [*Niebling et al,* 2010a, 2010b]. No other fluid effect was included. In particular, excess pore pressure above the hydrostatic pressure is not included, and flow of water between grains is assumed slow enough to neglect fluid viscosity effects. Density of particles is 1050 kg/m$^3$, corresponding to the experiments, and the interparticle friction coefficient is 0.6, producing a global sliding friction coefficient of μ~0.3 [*Morgan et al,* 2004], and negligible cohesion. The average diameter of the particles is 4 mm and 5% polydispersity, intruder diameter 24 mm, repulsive forces with spring constant 2 kN/m and viscosity corresponding to damping during particle interactions is 0.3 Pa.s. Shaking was imparted by oscillatory motion of sidewalls.

**5.2 Simulation results**

Figure 3 presents grain velocity snapshots from simulations at different $\Gamma$ and different water levels. Insets show the intruder position before and after shaking. The observed micromechanics agree with our theory: We used the same definition of liquefaction as for the experiments (see section 6).

Simulations demonstrate that liquefaction occurs when $\Gamma > \Gamma_L$ (Fig 3c, d), as expected from the Theory. In other cases (Fig 3 b, e) the intruder doesn't approach its isostatic depth. In the fully immersed case it is not clear how to define liquefaction since the initial state is nearly identical to the isostatic depth expected after liquefaction. Heterogeneous Liquefaction (Fig 3c) occurs for $\Gamma_{GE} > \Gamma > \Gamma_L$ and is associated with a static granular region under the intruder, where grains do not slide due to the high normal stress exerted by the intruder. Outside of this static region the granular medium rearranges, allowing sinking of the intruder. When $\Gamma > \Gamma_{GE}$ the whole medium slides, as predicted (Fig 3d). Fig 3f, g present the time evolution of the 2$^{nd}$ type of normalized emerged volume of an intruder

$$\Sigma_2 = \frac{V_{em}(t)-(V_{TOT}-V_{imISO})}{V_{em}(0)-(V_{TOT}-V_{imISO})}, \quad (4)$$

for these simulations and others and for the experiments presented in Fig 1. Here $V_{imISO}$ is the immersed volume expected at the isostatic position. The rest of the variable definitions are given in Table 1. $\Sigma_2$ starts at 1 and decays as the intruder sinks. For liquefied simulations (Fig 3f orange and brown curves) and experiments (Fig 3g orange curve), $\Sigma_2$ approaches 0, indicating that intruders approach their isostatic immersion, $V_{imISO}$. The fully immersed case is highly fluctuative since it starts very close to its isostatic depth.

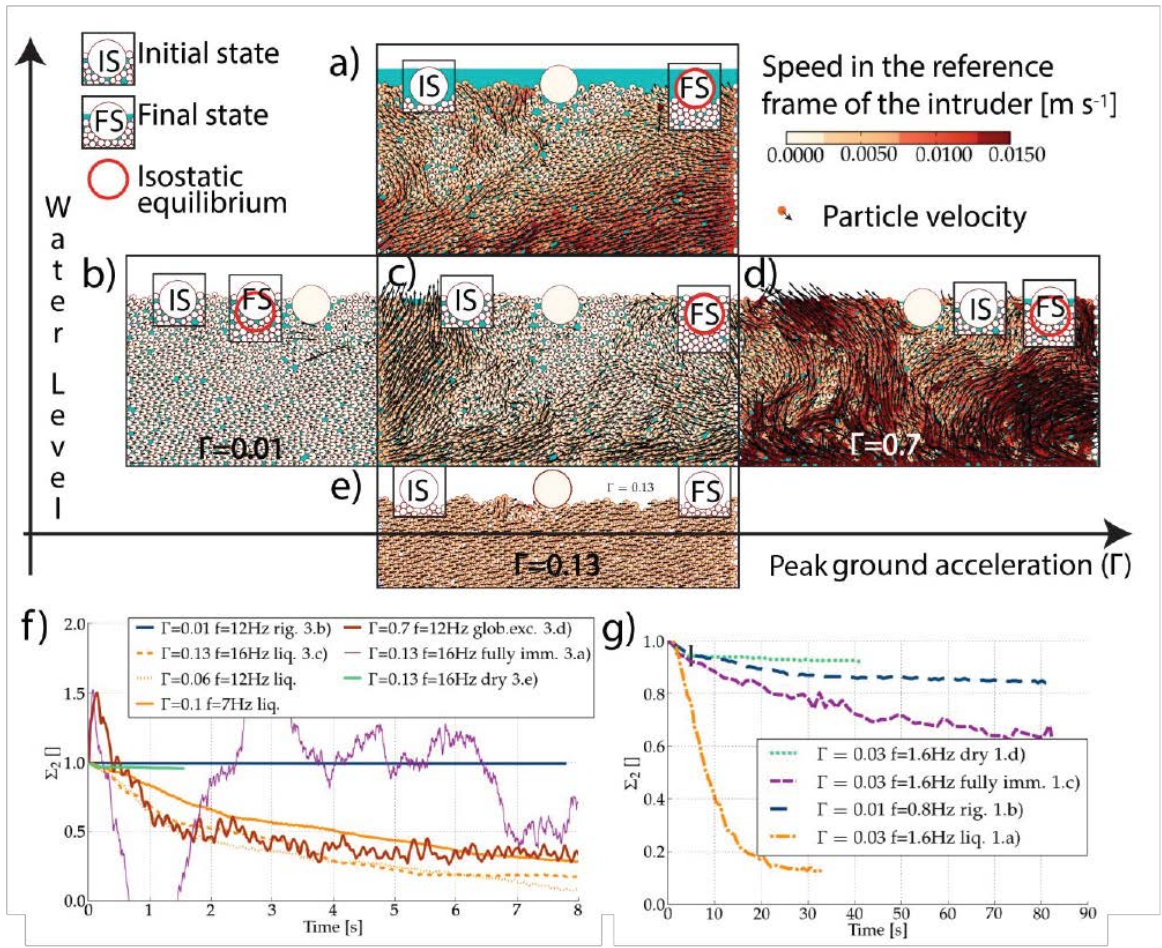

Fig 3: Snapshots from numerical simulations show that the micromechanics controlling liquefaction agrees with conceptual model of Fig 2. The velocity field is in the intruder reference frame. $\Gamma$ increases from left to right, and water level increases from bottom to top (snapshot e is dry). Insets present initial (I.S.) and final (F.S.) intruder

positions. Liquefaction occurs in two cases: Heterogeneous Liquefaction in (c), where grains far from the intruder slide but grains close to it dont slide, as seen on the velocity field, and Global Excitation in (d), where grains slip both far and close to the intruder. The fully immersed situation is too noisy to analyze. (f) Temporal evolution of Normalized emerged volume, $\Sigma_2$, for simulations. (g) Similar plot for the experiments shown in Fig 1. Simulation videos provided in Suppl. Movie S5 (b), Movie S6 (c), Movie S7 (d)

## 6 Analysis: Comparing Experiments, Simulations and Theory in a Phase diagram

Three main behaviors are observed in our experiments and simulations. The first behavior, termed *rigid*, is observed at low $\Gamma$. We define practically the medium as rigid if the medium rearrangements are negligible, i.e. if the 2$^{nd}$ type of normalized emerged volume of the intruder, $\Sigma_2$, (defined in equation 4) decreases from its initial value of 1 by less than 10% by the time it reaches its steady state position, i.e. $\Sigma_{2final}>0.9$. The second behavior, termed *Heterogeneous Liquefaction*, corresponds to monotonic subsidence of the intruder, where $\Sigma_{2final}<0.9$ but final oscillations remain small. The third behavior, termed *Globally Excited Liquefaction*, appears when the box is shaken with a high acceleration, and grains move both under the intruder and around it: in this case, the intruder sinks significantly as well ($\Sigma_2$ decreases below 0.9, the medium is liquefied), but does not reach a final equilibrium position, and oscillations of the intruder with respect to the surrounding medium persist at all times. Quantitatively, experiments and simulations are classified in this category when the standard deviation of the acceleration signal of the intruder exceeds an average value of 0.06 $g$.

Fig 4 presents a phase diagram showing the response of experimental and numerical systems as function of frequency and acceleration of shaking. Theory predicts that transitions between rigid deformation and liquefaction will be a straight line on this plot, depending only on acceleration. The predicted onset of liquefaction for our conditions, $\Gamma_L = 0.048$ μ, (transition from blue to orange symbols) is close to that observed in experimental and numerical simulations. Compared to field observations, our experiments and simulations used light beads to lower the value of $\Gamma_L$ and expand the range of accelerations allowing liquefaction. Since $\mu = 0.48$ for experiments, while $\mu = 0.3$ for simulations, a normalization of $\Gamma$ by $\mu$ is applied to plot them together.

Predictions for transition from heterogeneous liquefaction to *G.E.L.*, $\Gamma_{GE1}$ and $\Gamma_{GE2}$ (for simulations and experiments respectively), are calculated from equation (2), considering the dependence on the initial position of the intruder, see section 4.3. For simulations run in a saturated medium, the mechanical equilibrium that serves as the initial state displays an intruder half immersed (see Fig. 3), so $V_{im}(0)$ in equation (2), is roughly equal to $0.5 V_{TOT}$. In our experiments $V_{im}(0)$ varies between $0.05 V_{TOT}$ and $0.2 V_{TOT}$, depending on the exact amount of water and porosity. Thus $\Gamma_{GE1} = 0.5\mu$ for simulations and $\Gamma_{GE2} \in [0.8\mu;\ 0.95\mu]$ for experiments.

There is a good agreement in the phase diagram Fig.4, between the theoretically predicted transitions (lines) and the experimentally and numerically observed transitions (transitions are depicted by a change in symbol shape). The main noticeable discrepancy is that the experimentally observed transition between *H.L.* and *G.E.L.* is larger than predicted theoretically by $\Gamma_{GE2}$ – although the order of magnitude is respected. Note that good agreement is achieved with three totally independent methods.

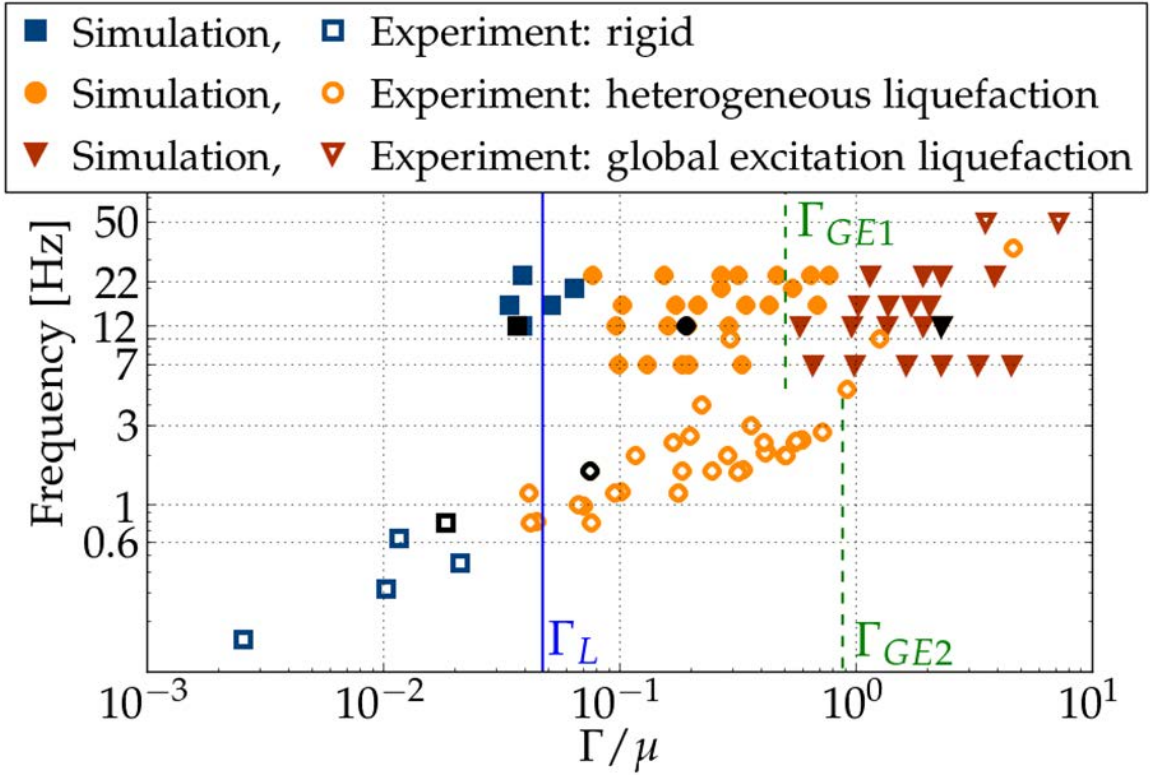

Fig 4: Phase diagram comparing simulations (filled symbols), experiments (open symbols) and theory (vertical lines predict transitions $\Gamma_L$, $\Gamma_{GE1}$ and $\Gamma_{GE2}$). Transitions in behavior observed in experiments and simulations (characterized according to definitions of section 6, and depicted by change in shape and color of symbols) agree well with theoretical predictions. Black symbols correspond to the experiments and simulations of Figs 1 and 3.

# 7 Discussion: Applicability of the new liquefaction mechanism to earthquakes.

## 7.1 Applicability of the new liquefaction mechanism to seismically shaken soils

The promising agreement between our theory, simulations and experiments (Fig 4), validates that liquefaction may be triggered by buoyancy plus acceleration effects alone. To check applicability of this scenario to nature, we next map our predicted acceleration conditions for liquefaction onset to seismically shaken soils. $\Gamma_L$ is material specific: For some types of clay friction is very low, e.g. in montmorillonite and illite µ = 0.2, $\rho_g$~2.65 kg/dm$^3$ [*Tembe et al,* 2010], leading to $\Gamma_{L\_clay} = 0.12$, while for sand µ =0.8 and $\rho_g$=2.65 kg/dm$^3$ [*Byerlee et al,* 1978] leading to $\Gamma_{L\_sand} = 0.5$. Fig 5 presents observations of international occurrences of liquefaction during earthquakes [*Wang,* 2007], showing that liquefaction occurs with seismic energy density as low as $e = 0.1$ J/m$^3$. Previous lab and field experiments find that liquefaction via the *conventional mechanism* requires $e > 30$ J/m$^3$ (dark grey region on picture) [*Wang* 2007, as interpreted from data from *Green et al,* 2004, *Liang et al* 1995, *Dief* 2000]. Thus the *conventional mechanism* explains only near-field liquefaction, which constitutes about half of observed liquefaction events [*Wang,* 2007; *Manga et al,* 2012], but fails to explain liquefaction events triggered in the far-field of earthquakes. To see if our process can explain those far-field liquefaction events, we translate our threshold acceleration $\Gamma_L$ to energy density. The calculation is detailed below.

Our aim here is to check which sites among the catalogue of the reference *Wang* [2007] were exposed to a $\Gamma$ beyond our predicted threshold $\Gamma_L = \mu(\rho_g - \rho_w)/\rho_g$. The normalized peak ground acceleration $\Gamma$ can be converted to energy density $e$ since $e = \rho_g v^2$ where v is the peak ground velocity. Hence for our mechanism the threshold energy for liquefaction onset is

$$e_L = \rho_g \left(\frac{g\Gamma_L}{\omega}\right)^2 \qquad (5)$$

To connect the magnitude $M_w$ of an earthquake to the energy density $e_L$ that it imparted at distance *r* from its hypocenter, we use the empirical relation accounting for the geometrical attenuation [*Wang,* 2007]:

$$M_w = 2.7 + 0.7\log_{10}(e_L) + 2.1\log_{10}(r) \qquad (6)$$

where $r$ is related to the epicenter distance $r_{surface}$ and the depth of the source $p$ through $r^2 = r^2_{surface} + p^2$. Hence we can plot the threshold energy density $e_L$ on the $(M_w, r)$ diagram on figure 5. To calculate $e_L$ using equation (5) for different earthquakes, we use the reference figure 8 from *Souriau* [2006] to obtain a characteristic ω via a relationship between the maximal dominant frequency $\omega/2\pi$ of earthquakes and the distance $r$:

$$\begin{cases} \omega/2\pi = 26 \text{ Hz for } \log(r) < 1.5 \\ \omega/2\pi = -20\log(r) + 60 \text{ Hz for } \log(r) > 1.5 \end{cases} \quad (7)$$

Given this relationship and the parameters μ and $\rho_g$ of a real soil for the computation of $\Gamma_L$, we obtain $e_L$ from equation (5) as a function of the source distance $r$. Next we plug $e_L$ in equation (6) and obtain:

$$M_w = 2.7 + 0.7\log_{10}\left(\rho_g({}^{g\mu}/_{2\pi})^2\right) + 1.4\log_{10}\left(\frac{\rho_g - \rho_w}{\rho_g}\right) - 1.4\log_{10}(26) + 2.1\log_{10}(r)$$

for $\log(r) < 1.5$ and **(8a)**

$$M_w = 2.7 + 0.7\log_{10}\left(\rho_g({}^{g\mu}/_{2\pi})^2\right) + 1.4\log_{10}\left(\frac{\rho_g - \rho_w}{\rho_g}\right) - 1.4\log_{10}(60 - 20\log(r))$$

$$+ 2.1\log_{10}(r)$$

for $\log(r) > 1.5$ **(8b)**

Our calculated limit for liquefaction onset is plotted on Figure 5, on a diagram of magnitude vs. epicenter distance as by *Wang* [2007]: The black lines 1 and 2 in Fig 5 represent equation (8), assuming $e_L$ is given by equation (5), and using $\Gamma_L = \mu\frac{\rho_g - \rho_w}{\rho_g} = 0.12$, with $\rho_g = 2.65$

kg/dm$^3$, μ = 0.2 and *g=9.8* m/s$^2$, representing very low friction clay-filled soils [*Byerlee,* 1978]. Every event in the light grey zone and below could be explained by our model.

The black lines are curved toward high magnitudes since dominant frequency of earthquakes decreases with distance from source [*Souriau,* 2006] (see equation (8)). We also checked that the offset between epicenter and hypocenter distance has a small impact on the position of our limit: the black line 2 takes this offset into account with an arbitrary depth source of 15 km for every event whereas the black line 1 assumes that all the events occur at the surface. The difference is only visible for low epicentral distance and can be neglected in our study. Our estimates for the furthest occurrences of $\Gamma$ = 0.12 in this representation, the black lines, based on equations (5-8), are directly confirmed by field measurements: the red points represent the furthest occurrence for $\Gamma$ = 0.12 recorded in a field example [*Wu et al,* 2001], and coincide reasonably well with the predicted black lines.

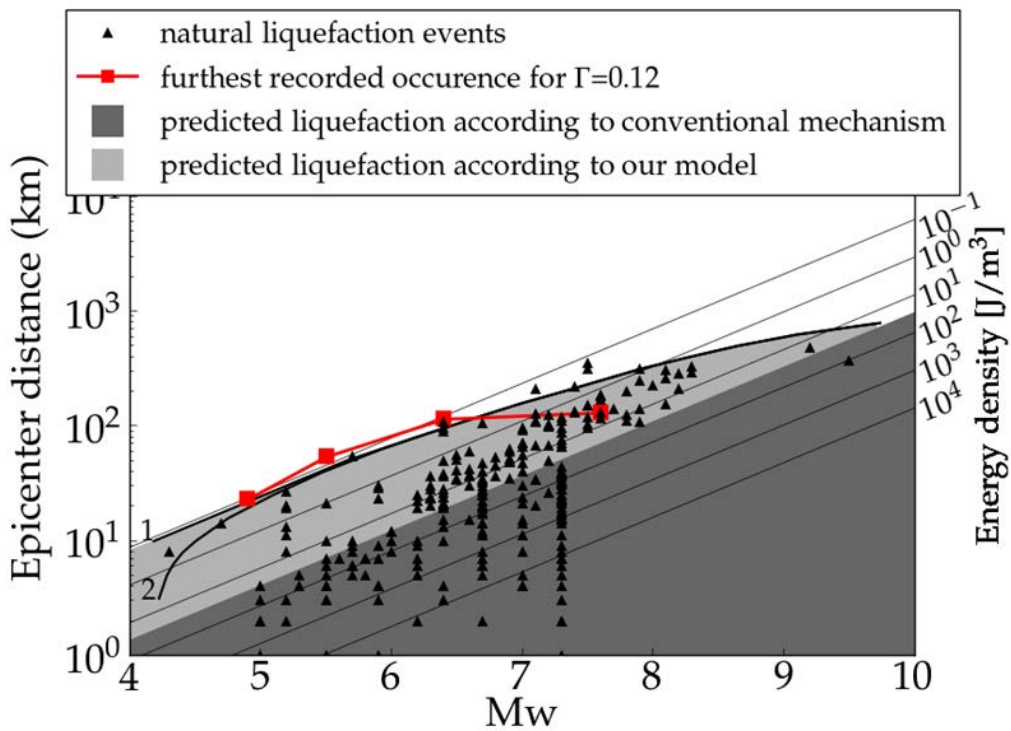

Manuscript submitted to *Journal of Geophysical Research B: Solid Earth*

Fig 5: Triangles show observations of liquefaction triggered by earthquakes of a given magnitude and epicentral distance (from *Wang* [2007], their Fig 1). The dark gray region depicts the regime of shaking conditions for which liquefaction is predicted to occur via the conventional mechanism, requiring energy density $e_L$> 30 J/m³. Only about a half of the observed earthquake-induced liquefaction events are predicted by the *conventional mechanism* for liquefaction. Our model predicts the previously unexplained far-field liquefaction region, and the shaking conditions allowing liquefaction by our model are depicted by the light grey region. Lines 1 &2 present our model theoretical limits for onset of liquefaction (equations (8)). The red points represent the furthest occurrence for $\Gamma$ = 0.12 recorded in a field example [*Wu et al,* 2001], confirming the estimates done in equations (5-8).

The limit for liquefaction plotted in Fig 5 (line 1&2) represents sites that are especially susceptible for liquefaction, as they liquefied despite the very low $\Gamma_L$ =*0.12* they experienced. Low acceleration can only liquefy soils with very low friction, as explained above, or soils with light grains or with somewhat elevated PP, as explained next: $\Gamma_L$ =*0.12* may alternatively represent soils with regular friction of $\mu = 0.6$, but with lighter grains, or mixtures of solid and trapped air, with an effective density for the grain/air clusters of only $\rho_g$ =*1.25$\rho_w$*, so that $\mu \frac{\rho_g - \rho_w}{\rho_g} = 0.12$. Eventually, an alternative scenario is that the pore pressure is elevated, exceeding the hydrostatic profile to reach 4/5 times the lithostatic stress (possibly due to upwards fluid motion and viscous forces, and/or due to compaction). Indeed, the normal force introduced in Section 4.1, leading to equation (1a), proportional to the effective stress, is in general $\vec{F}_{ij}^n = m_{above}(1 - P/\rho_g gH)\vec{g}$, while the tangential one is still $F_{ij}^t = A\omega^2 \, m_{above}$. The criterion $(F_{ij}^t / F_{ij}^n) > \mu$ can then be expressed as $\Gamma = \frac{A\omega^2}{g} > \Gamma_L = \mu(1 - P/\rho_g gH)$ : with an elevated pore pressure larger than hydrostatic $\rho_w gH$, and equal to 4/5 of the lithostatic stress, $P = 0.8\rho_g gH$, the limit becomes again $\Gamma_L = 0.2\mu = 0.12$.

Thus, shaking conditions close to our liquefaction limit (black) lines in Fig 5, may liquefy sites with low soil friction (such as clay) or sites with somewhat elevated PP, while conditions deeper in the grey region of Fig 5 can liquefy normal-friction soils with normal PP.

The above calculation shows that high pore pressure, generated via the conventional mechanism, or via pore pressure advection after breaking of permeability barriers [*Wang,* 2007; *Manga et al,* 2012; *Brodsky et al,* 2003], may enhance our proposed process, promote liquefaction onset and expand the light grey area on Fig 5. The red symbols correspond to the largest epicenter distance for events of measured $\Gamma = 0.12 = \Gamma_{L\_clay}$ for four ranges of magnitude in a catalogue [*Wu et al,* 2001]. These points confirm that our theoretical prediction for the relation between $\Gamma_L$ and earthquake distance and magnitude that produced it, is consistent with real shaking conditions.

The mechanism proposed here also explains the transition to sinking in shaken saturated quicksand, observed to happen in experiments at $\Gamma_{L\_quicksand} > 0.31$ [*Khaldoun et al,* 2005]. The experiments on quicksand in *Khaldoun et al.* [2005] used a mixture of 10% bentonite (montmorillonite) clay with 90% quartz sand, which *Tembe et al.* [2010], fig 4, measured to have $\mu = 0.6$. With that $\mu$, equation (1a) predicts the previously unexplained transition to sinking in quicksand will occur at $\Gamma_{L\_quicksand} = 0.37$, close to the $\Gamma_{L\_quicksand}$ observed in *Khaldoun et al.,* [2005]. In addition, sinking in quicksand is observed to proceed to isostatic depth [*Khaldoun et al.,* 2005] as predicted in the current theory. The current theory also agrees with the measured amplitude of the subsidence of buildings during liquefaction, as shown in the following paragraph.

**7.2 Final vertical position of a building resting on a liquefied soil dictated by our mechanism**

We check the applicability of this study for a typical concrete six-floor high building with superficial foundations of 1 meter below the surface, and total height $h_{TOT}$ =21 m. The density of such a building is simply its weight divided by its volume, estimated to be $\rho_B$ =0.39 kg/L$^3$, which is lower than water because a building is mostly composed of void. We run a set of experiments with an intruder of similar density in the polystyrene granular medium and observed that it sinks quickly to the final position dictated by isostasy. We next apply our model for real soils: For a soil of particle density $\rho_g = 2.7$ kg/$L^3$ and porosity of 0.4, $\rho_{eff} \sim 2$ kg/L$^3$ (see sec. 4.2). The expected sinking $h_{ISO}$ of this 6 floor building is then $h_{ISO} = h_{TOT}\ {\rho_B}/{\rho_{eff}} = 4$ m, which corresponds to 3m sinking, because the building has already foundation going 1m deep into the soil. Compared to field observations of sinking during liquefaction, our isostatic height has the right order of magnitude, although it is larger than most observations [*Soga,* 1998; *Bilham et al,* 2003; *Huang et al,* 2013]. The ground shaking duration during earthquakes is likely to be too short to allow buildings to reach their isostatic position. Indeed, the characteristic time for sinking is highly dependent on the size ratio between the object which sinks and the particles of the soil, according to our simulations. With a wider object more particles need to be involved and move away from under the structure to allow the descent of the object. In our experiments the half-life time for reaching isostatic equilibrium is tens of seconds to minutes. Because buildings are hundred times bigger and real soils have larger friction coefficient, several decades of minutes would likely be necessary for the building to reach half of its new equilibrium, but strong ground shaking is typically shorter. It is expected that the rate at which the 'building' sinks towards its isostatic equilibrium position is dictated by the balance of forces between the buoyancy forces of the medium on the intruder, and viscous resistive forces imposed by the granular medium, where the viscosity is non-trivial [*Sawicki et al,* 2009; *Cundall et al,* 1979; *Boyer et al,* 2011]. This dynamic approach to steady-state shall be left for future study.

## 8 Conclusions

In summary, theory, simulations and experiments combine to suggest a new liquefaction mechanism, that requires no fluid pressurization, and arises due to accelerations, friction and buoyancy in saturated granular media. Shaking of saturated granular media at an acceleration exceeding a *buoyancy–dictated threshold* (equation 1), causes granular sliding. Although pore pressure is not elevated in our experiments (see section 3.2), the sliding medium liquefies at a relatively low acceleration, allowing intruders to sink to isostatic depths dictated by buoyancy. Also the simulations, which included only buoyancy (and no excess pore pressure), show the same type of sinking, in the same predicted conditions, again proving that buoyancy forces alone promote liquefaction during shaking. Buoyancy-controlled-liquefaction may be applied to the Earth, to enlarge the window of conditions under which liquefaction is currently predicted to occur, and explain previously unexplained liquefaction cases (Fig 5). This mechanism is expected to operate in conjunction with the conventional mechanism of liquefaction. Although we show that the conventional mechanism and its associated high pore pressure are not necessary to produce liquefaction, and that buoyancy forces coupled with shaking are sufficient to induce liquefaction, if pore pressure is elevated, via the conventional mechanism or via another mechanism, (e.g. rupture of a previously sealed high pore pressure layer), it does enhance the process we present, promoting liquefaction at a lower acceleration. Buoyancy-controlled-liquefaction also explains the observed transition to sinking due to shaking in quicksand [*Khaldoun et al,* 2005], and the final sinking depth.

**Acknowledgments, Samples, and Data**

The authors declare no competing financial interests. We thank Alain Steyer and Miloud Talib for technical support, and Knut Jørgen Måløy, Ernesto Altshuler, Chris Scholz, Liran Goren, Gerhard Schäfer, Valérie Vidal, Mustapha Meghraoui Mohammed Bousmaha and Amir Sagy for interesting discussions. R.T. and E.A. acknowledge support of the European Union's Seventh Framework Programme for research, technological development and demonstration under grant agreement no 316889 (ITN FlowTrans), and of the CNRS INSU ALEAS program. The data for this paper are available by contacting the corresponding author at renaud.toussaint@unistra.fr

Manuscript submitted to *Journal of Geophysical Research B: Solid Earth*Green, R.A., Mitchell, J.K. (2004), Energy-based evaluation and remediation of liquefiable soils, Geotechnical Engineering for Transportation Projects. *ASCE Geotechnical Special Publication*, **2,** 1961-1970

Hausler, E. A., & Sitar, N. (2001), Performance of soil improvement techniques in earthquakes. *4th International Conference on Recent Advances in Geotechnical Earthquake Engineering and Soil Dynamics*, San Diego, USA.

Holzer , T., Hanks, T., Youd, T. (1989), Dynamics of liquefaction during the 1987 Superstition Hills, California, earthquake. *Science*, **244,** 56-59

Huang, Y. & Yu, M. (2013), Review of soil liquefaction characteristics during major earthquakes of the twenty-first century. *Natural hazards*, **65,** 2375-2384

Huang, Y., & Yu, M. (2013), Review of soil liquefaction characteristics during major earthquakes of the twenty-first century. *Natural hazards*, **65,** 2375-2384

Huerta, D.A., Sosa V., Vargas M.C., Ruiz-Suárez, J.C. (2005), Archimedes' principle in fluidized granular systems. *Phys. Rev. E*. **72,** 031307

Kawakami, F., & Asada, A. (1966), Damage to the ground and earth structures by the Niigata earthquake of June 16, 1964. *Soils and Foundations*, **6**, 14-30

Khaldoun, A., Eiser, E., Wegdam, G.H., Bonn, D. (2005), Rheology: Liquefaction of quicksand under stress. *Nature*, **437,** 635-635

Lakeland, D. L., Rechenmacher, A., Ghanem, R. (2014), Towards a complete model of soil liquefaction: the importance of fluid flow and grain motion. *Proceedings of the Royal Society of London A: Mathematical, Physical and Engineering Sciences*, **470,** 20130453
31

Table 1. List of symbols used

| Symbol | Description |
|---|---|
| $A$ | Amplitude of the displacement of the imposed external horizontal motion |
| $a$ | Peak ground acceleration : amplitude of the imposed external horizontal acceleration |
| $\beta$ | Fluid (water) compressibility |
| $D = k/\beta\eta\Phi$ | Pore Pressure diffusivity |
| $De = t_d/t_0$ | Deborah number |
| $e$ | Seismic energy density |
| $e_L$ | Seismic energy density at the onset of liquefaction |
| $\eta$ | Fluid (water) dynamic viscosity |
| $\vec{F}_{ij}^n$ | Vertical contact force exerted on bottom grain in the pair *ij* |
| $F_{ij}^t$ | Horizontal force exerted on bottom grain in the pair *ij* |
| $\vec{F}_{Bk}^n$ | Vertical contact force exerted on bottom grain in the pair *Bk* (under the intruder) |
| $\vec{F}_{Bk}^t$ | Horizontal contact force exerted on bottom grain in the pair *Bk* (under the intruder) |
| $f$ | Frequency of the imposed external horizontal acceleration |
| $\Phi$ | Soil (granular assembly) porosity |
| $g$ | Gravitational acceleration |
| G.E.L. | Globally Excited Liquefaction |
| $\Gamma$ | Dimensionless imposed acceleration amplitude, a/ g. |
| $\Gamma_{GE} = \mu$ | Limit in dimensionless acceleration between Heterogeneous Liquefaction and Globally Excited Liquefaction |
| $\Gamma_{GE1}$ | Limit in dimensionless acceleration between H.L. and G.E.L., taking into account the initial finite immersion of the intruder (in simulations) |
| $\Gamma_{GE2}$ | Limit in dimensionless acceleration between H.L. and G.E.L., taking into account the initial finite immersion of the intruder (in experiments) |
| $\Gamma_L = \mu(\rho_g - \rho_w)/\rho_g$ | Limit in dimensionless acceleration between rigid behavior and Heterogeneous Liquefaction |
| $H$ | Thickness of granular layer |
| $h_{TOT}$ | Height of building |
| $h_{ISO}$ | Isostatc depth (maximum expected sinking depth of building) |
| H.L. | Heterogeneous Liquefaction |
| $i$ | Index of grain considered in the medium |
| $j$ | Index of grain considered in the medium |
| $K$ | Soil (granular assembly) permeability |
| $k$ | Index of grain considered in the medium |
| $L$ | Thickness of dry granular zone on top of the saturated granular layer, or thickness of clear fluid above the top of the saturated granular layer |
| $l$ | Depth of liquefied layer |
| $m_{above}$ | Mass of grains in a column above a certain grain |
| $M_w$ | Earthquake magnitude |
| $\mu$ | Friction coefficient between grains |
| $p$ | Depth of hypocenter |
| PP | Pore pressure |
| $r$ | Distance to hypocenter |
| $r_{surface}$ | Distance to epicenter |
| $\rho_B$ | Density of the intruder |
| $\rho_g$ | Bulk density of the grains |
| $\rho_w$ | Density of the water |
| $\Sigma_1(t) = \dfrac{V_{em}(0) - V_{em}(t)}{V_{TOT}}$ | Normalized immersed volume |
| $\Sigma_2 = \dfrac{V_{em}(t) - (V_{TOT} - V_{imISO})}{V_{em}(0) - (V_{TOT} - V_{imISO})}$ | Normalized emerged volume |
| $t$ | Time after the start of the shaking |
| $t_0$ | Time to build pore pressure |
| $t_d$ | Time to diffuse pore pressure |
| $V_{em}$ | Emerged volume of the intruder |
| v | Peak ground velocity |
| $V_{em}(0)$ | Initial emerged volume of the intruder |
| $V_{em}(t)$ | Emerged volume of the intruder at time *t* |
| $V_{im}$ | Immersed volume of the intruder |
| $V_{im}(0)$ | Initial immersed volume of the intruder |
| $V_{imISO}$ | immersed volume expected at the isostatic position |
| $V_{TOT}$ | Total volume of the intruder |
| $\omega$ | Frequency of the imposed shaking |